\documentclass[11pt,a4paper]{article}
\usepackage{amsmath}
\usepackage{cite}
\usepackage{graphicx}
\usepackage{amsfonts}
\usepackage{amssymb}

\def\bt{
\beta}

\def\bt'{\beta'
}
\begin{document}
\bigskip
\hfill\hbox{SISSA 31/2004/FM} \vspace{2cm}

\begin{center}
{\Large \textbf{Form factors in the $SS$ model\\
\vspace{0.8cm}
and its RSOS restrictions}}\\
\vspace{1.2cm} {\Large B\'en\'edicte Ponsot \footnote{\textsf
{ponsot@fm.sissa.it}}} \\
\vspace{0.7cm} {\it  International School for Advanced Studies (SISSA),\\
Via Beirut 2-4, 34014 Trieste, Italy}\\
\vspace{1.2cm}
\end{center}

\begin{abstract}
New integral representations for form factors in the two
parametric $SS$ model are proposed. Some form factors in the
parafermionic sine-Gordon model and in an integrable perturbation
of $SU(2)$ coset conformal field theories are straightforwardly
obtained by different quantum group restrictions. Numerical checks
on the value of the central charge are performed.
\end{abstract}
\begin{center}
\end{center}
\section{Introduction}
Form factors with $n$ particles of an operator are the matrix
elements of this operator between the vacuum and a $n$ particles
state. They are non pertubative objects that can be constructed,
up to a normalization, as solution of "bootstrap" equations
\cite{KW,BKW,smirnov2}, once the exact scattering matrix is known.
Besides being solutions of a nice mathematical problem, the form
factors are useful tools to determine the long distance expansion
of correlation functions by inserting a complete set of asymptotic
states. It is often enough to approximate (with a good accuracy)
the correlation functions of local operators with the contribution
of the form factor with the smallest number of particles, due to
the fast convergence of the spectral series \footnote{It is
usually believed that the spectral
 series converges for local operators, but this has not been proven so far.}.

In this paper, we construct form factors in a two parametric
family of massive integrable quantum field theories known as the
$SS$ model, whose action can be found in \cite{fateev}, and which
can be written in terms of three boson fields $\varphi_i,
\;i=1,2,3$ with an exponential interaction. We will restrict
ourselves to the unitary regime of this theory \cite{fateev},
 where the model has a $U(1)\times U(1)$ symmetry described by two
 conserved topological charges:
\begin{eqnarray}
Q_{\pm}=\frac{1}{2}(Q_1\pm Q_2), \quad Q_i=\int dx^1 j_i^0, \quad
i=1,2\; , \nonumber
\end{eqnarray}
where $j_i^{0,1}$ are the components of the $U(1)$ current. Both
classically and in the quantum case, the charges (eigenvalues of
the charges) satisfy the conditions:
$$
Q_1,Q_2 \in \mathbb{Z}, \quad Q_1+Q_2 \in 2\mathbb{Z}.
$$
The $U(1)\times U(1)$ symmetry can be extended up to the symmetry
generated by
 two quantum affine algebras $U_{q_1}(\hat{sl_2})\otimes U_{q_2}(\hat{sl_2})$.
  The name of the $SS$ model is
  unfortunately justified by the expression of its two-particles $S$-matrix~-first
  considered
  in \cite{S}-~in terms of two sine-Gordon
  $S^{SG}$-matrices with different coupling constants $\beta_1,\beta_2$:
$$
S(\theta_{12})\equiv -S_{p_1}^{SG}(\theta_{12})\otimes
S_{p_2}^{SG}(\theta_{12}),
$$
where $p_1=\frac{\beta_1^2}{8\pi-\beta_1^2},\;
p_2=\frac{\beta_2^2}{8\pi-\beta_2^2}$  \cite{ZZ}, and
$\theta_{12}=\theta_1-\theta_2$ is the rapidity difference. The
unitary regime is characterized by the following conditions on the
parameters: $p_1,p_2>0$ and $p_1+p_2\geq 2$. We will need to introduce the negative parameter
$p_3$ such that $p_1+p_2+p_3=2$,
 and the parameters $\alpha_i$, $i=1,2,3$, defined by $p_i=2\alpha_i^2$ \cite{fateev}.\\
The $SS$ model includes,
as particular cases, the N=2 supersymmetric sine-Gordon model,
whose $S$-matrix has the form \cite{ku}:
$$
S(\theta_{12})= -S_{2}^{SG}(\theta_{12})\otimes
S_{p_2}^{SG}(\theta_{12}),
$$
as well as the $O(4)$ non linear sigma model (Principal chiral
field model \cite{PW,FR}), which possesses $SU(2)\times SU(2)$
symmetry; its $S$-matrix reads \cite{ZZ}
$$
S(\theta_{12})=-S_{\infty}^{SG}(\theta_{12})\otimes
S_{\infty}^{SG}(\theta_{12}),
$$
which is nothing but the tensor product of two $S$-matrices of the
$SU(2)$ invariant Thirring model.
 It also contains the anisotropic
chiral field \cite{W}, which has the $U(1)\times SU(2)$ symmetry;
its $S$-matrix is
$$
S(\theta_{12})= -S_{p_1}^{SG}(\theta_{12})\otimes
S_{\infty}^{SG}(\theta_{12}).
$$
The $SS$ model also includes other known integrable QFTs, like the
$O(3)$ non linear sigma model, the
sausage model \cite{FOZ} and the cosine-cosine model \cite{BL}.
Let us note in passing that the ground state energy of the model in
finite volume was determined in \cite{fateev} by TBA method, and the finite size
 effects for the $SS$ model were recently studied in \cite{H}.\\
In order to construct the form factors in this theory, we give in
the first section a summary of the method used in \cite{BFKZ,BK}
to construct form factors in the sine-Gordon model, and recall how
to perform RSOS restriction \cite{L-RS} directly on sine-Gordon
form factors \cite{BBS}. In the second section, we propose a
generic formula inspired from the SG model for form factors
containing an even number of particles in the $SS$ model; in
particular we write down the form factors of the trace of the
energy momentum tensor, first obtained in \cite{S}, and of some of
the exponential fields (this problem was also recently considered
in \cite{FL}, with different methods). Two particles form factors
of these operators can be expressed in terms of two particles form
factors in the SG model. Then we do two different kinds of RSOS
restrictions, namely \cite{ABL} to the parafermionic sine-Gordon
model (this model includes, amongst others, the N=1 and the
restricted N=2 supersymmetric sine-Gordon models), and to the
integrable perturbed coset CFT \cite{GKO} $su(2)_{p_1-2} \otimes
su(2)_{p_2-2} /su(2)_{p_1+p_2-4}$. A common limit of these QFT's
is the Polyakov-Wiegmann model \cite{PW}. We propose expressions
for form factors of the trace operator (they were first considered
in \cite{S}). In particular, we check our expressions for the two
particles form factor of the trace of the energy momentum tensor
by making a numerical estimation of the central charge and compare
it with the exact result.

\section{Form factors in the sine-Gordon model and RSOS restriction}
In this section we recapitulate known results on form factors in
the SG model in the repulsive regime. They will be useful in the
next section for the construction of the form factors in the
$SS$ model.\\
The Sine-Gordon model alias the massive Thirring model is defined
by the Lagrangians:
\begin{eqnarray}
\mathcal{L}^{SG}&=&\frac{1}{2}(\partial_{\mu}\varphi)^2+\frac{\alpha}{\beta^2}(\cos \beta\varphi-1),\label{lag}\\
\mathcal{L}^{MTM}&=&\bar{\psi}(\mathrm{i}\gamma\partial-M)\psi-\frac{1}{2}g(\bar{\psi}\gamma_{\mu}\psi)^2,\nonumber
\end{eqnarray}
respectively. The fermi field $\psi$ correspond to the soliton and
antisoliton and the bose field $\varphi$ to the lowest breather
which is the lowest soliton antisoliton bound state.
 The relation between the coupling constants was found in \cite{Coleman}
 within the framework of perturbation theory:
$$
p\equiv\frac{\beta^2}{8\pi-\beta^2}=\frac{\pi}{\pi+2g}.
$$
The two soliton sine-Gordon $S$-matrix contains the following
scattering amplitudes: the two-soliton amplitude $a_p(\theta)$,
the forward and backward soliton anti-soliton amplitudes
$b_p(\theta)$ and $c_p(\theta)$:
\begin{eqnarray}
b_p(\theta)= \frac{\sinh\theta/p}{\sinh(i\pi-\theta)/p}a_p(\theta),
\quad c_p(\theta)=\frac{\sinh i\pi/p } {\sinh (i\pi-\theta)/p
}a_p(\theta),\nonumber
\end{eqnarray}
\begin{eqnarray}
a_p(\theta)=\mathrm{exp}\;
\int_{0}^{\infty}\frac{dt}{t}\frac{\sinh\frac{1}{2}(1-p) t \sinh
\frac{t\theta}{i\pi}}{\cosh \frac{t}{2}\sinh\frac{1}{2}p
t}.\nonumber
\end{eqnarray}
This $S$-matrix satisfies the Yang-Baxter equation as well as the
unitarity condition:
$$
S_p^{SG}(\theta)S_p^{SG}(-\theta)=1,
$$
which can be rewritten for the amplitudes as:
$$
a_p(\theta)a_p(-\theta)=1,\quad
b_p(\theta)b_p(-\theta)+c_p(\theta)c_p(-\theta)=1.
$$
The crossing symmetry condition reads for the amplitudes:
$$
a_p(\mathrm{i}\pi-\theta)=b_p(\theta),\quad
c_p(\mathrm{i}\pi-\theta)=c_p(\theta).
$$
The form factors $f(\theta_1,\cdots,\theta_{2n})$ of a local
operator\footnote{We shall consider
 in the following form factors with an even number of particles only.} in the SG
model are covector valued functions that satisfy a system of equations \cite{smirnov2}, which consist
of a Riemann-Hilbert problem:
\begin{eqnarray}
&&f(\theta_1,\cdots,\theta_{i},
\theta_{i+1},\cdots,\theta_{2n})S_p^{SG}(\theta_i-\theta_{i+1})=
f(\theta_1,\cdots,\theta_{i+1},
\theta_i,\cdots,\theta_{2n}),\nonumber \\
&& f(\theta_1,\cdots,\theta_{2n-1},\theta_{2n}+2i\pi)=
 f(\theta_{2n},\theta_1,\cdots,\theta_{2n-1}),\nonumber
\end{eqnarray}
and a residue equation at $\theta_1=\theta_{2n}+i\pi$ :
\begin{eqnarray}
\mathrm{res}f(\theta_1,\cdots,\theta_{2n})=-2i\;
f(\theta_2,\cdots,\theta_{2n-1})
\left(1-\prod_{i=2}^{2n-1}S_p^{SG}(\theta_{i}-\theta_{2n})\right)
e_0,\quad e_0=s_{1}\otimes \bar{s}_{2n} + \bar{s}_1\otimes s_{2n},
\nonumber
\end{eqnarray}
where $s$ (or +) corresponds to the solitonic state (highest
weight state), and $\bar{s}$ (or -) to the antisolitonic state.
Form factors containing an arbitrary number of particles
for the energy momentum tensor, the topological current and the
semi-local operator $e^{\pm i\frac{\beta}{2}\varphi_{SG}}$ were
first constructed in \cite{smirnov2} by Smirnov; form factors of
non local exponential fields $e^{ia\varphi_{SG}}$ were constructed
in\cite{L} by Lukyanov, using free field representation techniques
that provide integral representations different from those of
\cite{smirnov2}. Below, we make the choice to present the
posterior construction
presented in \cite{BFKZ,BK} by Babujian and Karowski {\it et. al.}\\
We first introduce the minimal form factor $f_p(\theta_{12})$ of
the SG model:
 it satisfies the relation
  $$f_p(\theta)=-f_p(-\theta)a_p(\theta)=f_p(2i\pi-\theta),$$
and reads explicitly
\begin{eqnarray}
f_p(\theta)=-i\sinh \frac{\theta}{2} f^{min}_p(\theta)=-i\sinh
\frac{\theta}{2}\exp
\int_{0}^{\infty}\frac{dt}{t}\frac{\sinh\frac{1}{2}(1-p)t}{\sinh\frac{1}{2}p
t\cosh\frac{1}{2}t} \frac{1-\cosh t(1-\frac{\theta}{i\pi})}{2\sinh
t}. \nonumber
\end{eqnarray}
Its asymptotic behaviour when $\theta \to \pm \infty$ is given by
$ f_p(\theta)\sim \mathcal{C}_p\;
e^{\pm\frac{1}{4}(\frac{1}{p}+1)(\theta-i\pi)} $, with the
constant
\begin{eqnarray}
\mathcal{C}_p=\frac{1}{2}\exp
\frac{1}{2}\int_{0}^{\infty}\frac{dt}{t}
\left(\frac{\sinh\frac{1}{2}(1-p)t}{\sinh\frac{1}{2}p
t\cosh\frac{1}{2}t\sinh t}-\frac{1-p}{p t}\right).
\end{eqnarray}
It is proposed in \cite{BFKZ} that form factors in SG can be
generically written\footnote{This representation holds whether operators are
 local or not, topologically neutral or not.}:
\begin{eqnarray}
f(\theta_1,\theta_2,\dots,\theta_{n})=N_{n}
\prod_{i<j}f_p(\theta_{ij})\int_{C_{\theta}}du_1 \dots
\int_{C_{\theta}}du_m\; h_p(\theta,u)
 p_n(\theta,u)\Psi^{p}(\theta,{u}),
\label{ff}
\end{eqnarray}
where we introduced the scalar function (completely determined by
the $S$-matrix)
\begin{eqnarray}
h_p(\theta,u)=\prod_{i=1}^{2n}\prod_{j=1}^{m}\phi_p(\theta_{i}-u_j)
\prod_{1\le r<s\le m}\tau_p(u_r-u_s), \nonumber
\end{eqnarray}
with
$$
\phi_p(u)=\frac{1}{f_p(u)f_p(u+i\pi)},\quad
\tau_p(u)=\frac{1}{\phi_p(u)\phi_p(-u)}.
$$
$\Psi^p(\theta,{u})$ is
the Bethe ansatz state covector: we first define the monodromy matrix $T_p$ as
\begin{displaymath}
\left(\begin{array}{cc}
A(\theta_1,\dots,\theta_n,u) & B(\theta_1,\dots,\theta_n,u)\\
C(\theta_1,\dots,\theta_n,u) & D(\theta_1,\dots,\theta_n,u)\\
\end{array}\right) \equiv
T_p(\theta_1,\dots,\theta_n,u)
= S^{SG}_p(\theta_1-u)\dots S^{SG}_p(\theta_{n}-u),
\end{displaymath}
the definition of the Bethe ansatz covector is given by
$$
\Psi^p(\theta,{u})= \Omega_{1\dots n}\;
C(\theta_1,\dots,\theta_n,u_1)\dots C(\theta_1,\dots,\theta_n,u_m),
$$
where $\Omega_{1\dots n}$ is the pseudo vacuum consisting only of
solitons
$$
\Omega_{1\dots n} = s\otimes\dots \otimes s.
$$
The number of integration variables $m$ is related to the
topological charge $Q$ ($Q\in \mathbb{Z}$) of the operator
considered and the number $n$
 of particles through the relation $Q=n-2m$. For example, for $n=2$ and $Q=0$,
$$
\Psi^p(\theta_1,\theta_2,u)=\Psi^p_{+-}(\theta_1,\theta_2,u)+\Psi^p_{-+}(\theta_1,\theta_2,u)=
b(\theta_1-u)c(\theta_2-u)s_1\otimes\bar{s}_2+c(\theta_1-u)a(\theta_2-u)\bar{s}_1\otimes s_2.
$$
 The function $p_{n}(\theta,u)$ is the only
ingredient in formula (\ref{ff}) which depends on the operator
considered. If the operator is chargeless, the form factors contain an even number of particles, and if in addition the operator is local,
 then the $p$-function satisfies the conditions\footnote{We consider here only the case where the operator is
 of bosonic type and the particles are of fermionic type. If both are fermionic, there is an extra
  statistic factor to be taken into account, see \cite{BK}.}:
\begin{enumerate}
\item
 $p_{2n}(\theta,u)$ is a polynomial in $e^{\pm
u_j}$, $(j=1,\dots ,n)$ and $p_{2n}(\theta,u)=
p_{2n}(\dots,\theta_i-2i\pi,\dots,u)$
\item
$p_{2n}(\theta_1=\theta_{2n}+i\pi,\dots\theta_{2n};u_1\dots u_n=\theta_{2n})=
p_{2n-2}(\theta_2\dots\theta_{2n-1};u_1\dots
u_{n-1})+\tilde{p}^{1}(\theta_2\dots\theta_{2n-1}),
$\\
$p_{2n}(\theta_1=\theta_{2n}+i\pi,\dots\theta_{2n};u_1\dots u_n=\theta_{2n}\pm i\pi)=
p_{2n-2}(\theta_2\dots\theta_{2n-1};u_1\dots
u_{n-1})+\tilde{p}^{2}_{\pm}(\theta_2 \dots \theta_{2n-1}), $\\ where
$\tilde{p}^{1,2}(\theta_2 \dots \theta_{2n-1})$ are independent of
the integration variables.
\item
$p_{2n}(\theta,u)$ is symmetric with respect to the $\theta$'s and
the $u$'s.
\item
$p_{2n}(\theta+\ln\Lambda,u+\ln\Lambda)=\Lambda^{\mathrm{s}}\;p_{2n}(\theta,u)$
where s is the Lorentz spin of the operator.
\end{enumerate}
Finally, the integration
contours $C_{\theta}$ consist of several pieces for all
integration variables $u_j$~: a line from $-\infty$ to $\infty$
 avoiding all poles such that
$\mathrm{Im}\theta_i-\pi-\epsilon<\mathrm{Im}
u_j<\mathrm{Im}\theta_i-\pi,$ and clockwise oriented circles
around the poles (of the $\phi(\theta_i-u_j)$) at $\theta_i=u_j$,
$(j=1,\dots,m)$.
\paragraph{\it Trace of the energy momentum tensor.}
The trace operator is a spinless and chargeless local operator.
 Its $p$-function is \cite{BK}:
\begin{eqnarray}
p_{SG}^{\Theta}(\theta,u)=-\left(\sum_{i=1}^{2n}e^{-\theta_i}\sum_{j=1}^{n}e^{u_j}-
 \sum_{i=1}^{2n}e^{\theta_i}\sum_{j=1}^{n}e^{-u_j}\right).
 \label{trace}
 \end{eqnarray}
The residue equation gives the following relation for the
normalization $N_{2n}^{\Theta}$ (see \cite{BFKZ}):
$$
N_{2n}^{\Theta}=N_{2n-2}^{\Theta}\frac{\left(f^{min}_p(0)\right)^2}{4n\pi}\quad
\to N_{2n}^{\Theta}=
N_{2}^{\Theta}\frac{1}{n!}\left(\frac{\left(f^{min}_p(0)\right)^2}{4\pi}\right)^{n-1}.
$$
The two particles form factor can be computed
explicitly:
\begin{eqnarray}
f_{SG}^{\Theta}(\theta_{12})= \frac{2\pi
N_2^{\Theta}}{\mathcal{C}_p^4}f_p(\theta_{12})\frac{\cosh
\frac{\theta_{12}}{2}}{\sinh\frac{1}{2p}(i\pi-\theta_{12})}(s_1\otimes
\bar{s}_2+\bar{s}_1\otimes s_2),
\nonumber
\end{eqnarray}
in agreement with the result first obtained by
diagonalization of the $S^{SG}$-matrix in \cite{KW}.
The normalization for two particles is chosen to be
$N_2^{\Theta}=\frac{1}{p}iM^2\mathcal{C}_p^4$ ($M$ being the mass of
the soliton), in order to have:
$$
f_{SG}^{\Theta}(\theta_1+i\pi,\theta_1)= 2\pi M^2 (s_1\otimes
\bar{s}_2+\bar{s}_1\otimes s_2),
$$

\paragraph{\it Exponential fields.} The $p$-function of the
spinless, chargeless, non local, exponential field
$e^{ia\varphi_{SG}(x)}$ in the SG model is not
   written in \cite{BK}, though it is a minor extension of the results
    contained in this paper. It reads
\begin{eqnarray}
p_{SG}^{\frac{a}{\beta}}(\theta,u)= \frac{1}{e^{\frac{i\pi a}{\beta}}}
\frac{\prod_{j=1}^{n} e^{\frac{2a u_j}{\beta}}}{\prod_{i=1}^{2n}
e^{\frac{a\theta_i}{\beta}}}. \label{exp}
\end{eqnarray}
The conditions 1. and 2. are modified into:
\begin{enumerate}
\item
$p^{\frac{a}{\beta}}_{SG}(\theta,u)= e^{-\frac{2i\pi a
}{\beta}}p^{\frac{a}{\beta}}_{SG}(\dots,\theta_i-2i\pi,\dots,u)$,
\item
$p^{\frac{a}{\beta}}_{SG}(\theta_1=\theta_{2n}+i\pi,\dots\theta_{2n};u_1\dots u_n=\theta_{2n})=
 e^{\frac{-2i\pi a
}{\beta}} p^{\frac{a}{\beta}}_{SG}(\theta_2\dots\theta_{2n-1};u_1\dots
u_{n-1}),\nonumber
$\\
$p^{\frac{a}{\beta}}_{SG}(\theta_1=\theta_{2n}+i\pi,\dots\theta_{2n};u_1\dots u_n=\theta_{2n}+i\pi)=
p^{\frac{a}{\beta}}_{SG}(\theta_2\dots\theta_{2n-1};u_1\dots
u_{n-1}), $\\
$p^{\frac{a}{\beta}}_{SG}(\theta_1=\theta_{2n}+i\pi,\dots\theta_{2n};u_1\dots
u_n=\theta_{2n}-i\pi)= e^{\frac{-4i\pi a
}{\beta}}p^{\frac{a}{\beta}}_{SG}(\theta_2\dots\theta_{2n-1};u_1\dots
u_{n-1}). $\\
\end{enumerate}
The form factors of exponential fields were first constructed in
\cite{L} with a different representation. For $a =
\frac{k}{2}\beta$, with $k\in \mathbb{Z}$, the form factors can be
computed explicitly, and their expression can be found in
\cite{L}; In particular, the form factors of the semi-local
operator $e^{\pm i\frac{\beta}{2}\varphi_{SG}(x)}$ were first
obtained in \cite{smirnov2}; their expression with two particles
 is\footnote{We will not need the normalization factor of the
 exponential fields in the SG model. They can be found in \cite{LZ}.}:
\begin{eqnarray}
f_{SG}^{\pm \frac{1}{2}}(\theta_{12})= f_p(\theta_{12})
\frac{i\pi}{\mathcal{C}_p^4\sinh \frac{1}{p}(i\pi
-\theta_{12})}\left(e^{\pm \frac{i\pi -\theta_{12}}{2p}}s_1\otimes
\bar{s}_2+e^{\mp\frac{i\pi -\theta_{12}}{2p}}\bar{s}_1\otimes
s_2\right).
\nonumber
\end{eqnarray}
An important remark to be made is that there exists an alternative
$p^{\Theta}$-function to eq.~(\ref{trace}) for the trace of the energy momentum tensor.
Indeed, if one remembers that in the SG model the trace $\Theta$ is identified as the
term in the action $\cos \beta\varphi_{SG}$, up to inessential coefficients,
then one can rewrite its $p$-function in a suggestive form as a sum of $p$-functions
for exponential fields with $a=\beta$ and  $a=-\beta$, {\it i.e.}:
\begin{eqnarray}
p_{SG}^{\Theta}(\theta,u)=p_{SG}^{1}(\theta,u)+p_{SG}^{-1}(\theta,u).
\label{altern}
\end{eqnarray}
Indeed, using the expression
\begin{eqnarray}
f_{SG}^{\pm 1}(\theta_{12})= f_p(\theta_{12})
\frac{2\pi\;\mathrm{cotan} \frac{\pi
p}{2}}{\mathcal{C}_p^4}\frac{\cosh\frac{\theta_{12}}{2}}{\sinh
\frac{1}{p}(i\pi -\theta_{12})}\left(e^{\pm \frac{i\pi
-\theta_{12}}{2p}}s_1\otimes \bar{s}_2+e^{\mp\frac{i\pi
-\theta_{12}}{2p}}\bar{s}_1\otimes s_2\right), \nonumber
\end{eqnarray}
we find
\begin{eqnarray}
f_{SG}^{\Theta}(\theta_{12})= \frac{2\pi
\tilde{N}_2^{\Theta}\mathrm{cotan} \frac{\pi
p}{2}}{\mathcal{C}_p^4}f_p(\theta_{12})\frac{\cosh
\frac{\theta_{12}}{2}}{\sinh\frac{1}{2p}(i\pi-\theta_{12})}(s_1\otimes
\bar{s}_2+\bar{s}_1\otimes s_2), \nonumber
\end{eqnarray}
where the new normalization constant $\tilde{N}_2^{\Theta}$ is
given this time by the formula
$\tilde{N}_2^{\Theta}=\frac{iM^2\mathcal{C}_p^4}{p}\; \tan
\frac{\pi p}{2}$.\\
Let us note that the vacuum expectation value of the trace was
obtained in \cite{FSZ} thanks to the thermodynamic Bethe ansatz,
and is equal to $<\Theta>=-\pi M^2 \tan \frac{\pi p}{2}$; the
following relation then holds:
$$
\tilde{N}_2^{\Theta}=-<\Theta>\frac{\mathcal{C}_p^4}{p\;\pi}.
$$
\paragraph{\it RSOS restriction \cite{L-RS}.}
The RSOS restriction describes the $\Phi_{1,3}$-perturbations
 of minimal models of CFT \cite{Z} for rational values of $p$. When $p$ is an integer, we deal with
$\Phi_{1,3}$-perturbations
 of minimal models $M_p$ with central charge $c=1-\frac{6}{p(p+1)}$.
In particular, we remind that $S_{3}^{RSOS}=-1$ is the Ising
$S$-matrix. Form factors in the model $M_p$ can be
 directly obtained from those of the SG model, as explained in
 \cite{BBS}. For the trace operator, the RSOS procedure consists
 of 'taking the half' of the $p$-function\footnote{The rationale behind this is explained in \cite{BBS}.}
 (\ref{trace}), such that its $p$-function reads:
\begin{eqnarray}
p_{RSOS}^{\Theta}(\theta,u)=-\sum_{i=1}^{2n}e^{-\theta_i}\sum_{m=1}^{n}e^{u_m},
\nonumber
\end{eqnarray}
then we should modify the Bethe ansatz state:
\begin{eqnarray}
\tilde{\Psi}^{p}_{\epsilon_1 \epsilon_2 \dots \epsilon_{2n}}\equiv
e^{\frac{1}{2p}\sum_i \epsilon_i \theta_i} \Psi^{p}_{\epsilon_1
\epsilon_2 \dots \epsilon_{2n}}, \quad \epsilon_i=\pm,\; \sum_{i=1}^{2n}\epsilon_i=0.
\label{ms}
\end{eqnarray}
The two particles form factor of the trace operator reads
explicitly:
\begin{eqnarray}
f_{RSOS}^{\Theta}(\theta_{12})= \frac{2\pi
N_2^{\Theta}}{\mathcal{C}_p^4}f_p(\theta_{12})\frac{\cosh
\frac{\theta_{12}}{2}}{\sinh\frac{1}{p}(i\pi-\theta_{12})}(e^{\frac{i\pi}{2p}}s_1\otimes
\bar{s}_2 + e^{-\frac{i\pi}{2p}}\bar{s}_1\otimes
s_2)\label{fthetarsos}
\end{eqnarray}
and the normalization for two particles is chosen to be
$N_2^{\Theta}=\frac{2}{p}iM^2\mathcal{C}_p^4$, in order to have:
$$
f_{RSOS}^{\Theta}(\theta_1+i\pi,\theta_1)= 2\pi
M^2(e^{\frac{i\pi}{2p}}\; s_1\otimes \bar{s}_2 +
e^{-\frac{i\pi}{2p}}\; \bar{s}_1\otimes s_2).
$$
\begin{center}
\begin{table}
\begin{center}
\begin{tabular}{|c|c|c|c|}
\hline
 $p$ & $c^{(2)}_{\mathrm{num}}$ & $c_{\mathrm{exact}}$\\
\hline
\hline
3&0.5&0.5\\
3.12&0.5331&0.533234\\
4&0.6988&0.7\\
5&0.7972&0.8\\
10&0.9373&0.9454...\\
20&0.9744&0.9857...\\
100&0.9864&0.9994...\\
\hline
\end{tabular}
\end{center} \caption{Minimal models $M_p$} \label{tab1}
\end{table}
\end{center}
The knowledge of the form factors of the trace of the stress
energy tensor allows to estimate the variation of the central
charge by means of the so-called "$c$-theorem" sum rule
\cite{Zamolodchikov:1986gt,Cardy:1988tj}:
\begin{eqnarray}
c_{UV}=\frac{3}{2}\int_{0}^{\infty}dr\;
r^3<\Theta(r)\Theta(0)>.\label{cth}
\end{eqnarray}
Since in the massive case any correlation function can be
represented by its spectral expansion
\begin{eqnarray}
\lefteqn{<O(x)O(0)>=} \label{int} \\
&& \sum_{n=0}^{\infty}\frac{1}{n!}\int_{-\infty}^{+\infty}
\frac{d\theta_1\dots d\theta_n}{(2\pi)^{n}}
|F_{n}(\theta_1,\ldots,\theta_{n})|^2
e^{-Mr\sum_{j=1}^{n}\cosh{\theta}_j }, \nonumber
\end{eqnarray}
the computation of $ c $ turns out to be a non trivial check for
the form factor $f^{\Theta}_{RSOS}(\theta_{12})$ in equation
(\ref{fthetarsos}). In Table 1 above are presented the numerical
results with two particles contribution for the central charge in
the $M_p$ model versus the theoretical result. As one may observe
with the case $p=3.12$, the parameter $p$ can be taken continuous,
as the observables depend continuously on it \footnote{We refer
the reader to the Fig.5 of \cite{Delfino} for similar numerical
tests on the central charge (the parameter $p$ is taken
continuous); in this article, the author constructed form factors
with two kinks only, starting directly with the definition of the
RSOS matrix. It is trivial to see
that our correlation function $<\Theta\Theta>$ with two particles coincides with the one of \cite{Delfino}.}.\\
More generally \cite{BBS}, one can obtain the form factors of the
primaries $\Phi_{1,k}$ using the identification
 $\Phi_{1,k}\sim e^{i\frac{(k-1)}{2}\beta \varphi_{SG}(x)}$, and then
'twisting' the Bethe ansatz state like in (\ref{ms}). In the two
particle case, the form factors can be computed explicitly.

\section{Form factors in the $SS$ model and its RSOS restrictions}
This program (for $p_1,p_2\geq 1$) was initiated for this model
first by Smirnov \cite{S} for form factors of the energy momentum
tensor and the $U(1)\times U(1)$ current, then continued recently
by Fateev and Lashkevich \cite{FL}, who proposed expressions for
the form
factors of the exponential fields using the method of \cite{L}. \\
The form-factors $F(\theta_1,\cdots,\theta_{2n})$ of a local
operator in the $SS$ model satisfy the system of equations
\cite{S}:
\begin{eqnarray}
&&F(\theta_1,\cdots,\theta_{i},
\theta_{i+1},\cdots,\theta_{2n})S(\theta_i-\theta_{i+1})=F(\theta_1,\cdots,\theta_{i+1},
\theta_i,\cdots,\theta_{2n}),\nonumber \\
&& F(\theta_1,\cdots,\theta_{2n-1},\theta_{2n}+2i\pi)=
 -F(\theta_{2n},\theta_1,\cdots,\theta_{2n-1}),\label{rh}
\end{eqnarray}
and the residue equation at $\theta_1=\theta_{2n}+i\pi$:
\begin{eqnarray}
\mathrm{res}F(\theta_1,\cdots,\theta_{2n})=
-2i\;F(\theta_2,\cdots,\theta_{2n-1})
\left(1-\prod_{i=2}^{2n-1}S_{p_1}(\theta_{i}-\theta_{2n})
S_{p_2}(\theta_{i}-\theta_{2n})\right) e_0,
 \label{res}
\end{eqnarray}
$e_0=(s_1\otimes \bar{s}_2 + \bar{s}_1\otimes
s_2)\otimes(s_1\otimes \bar{s}_2 + \bar{s}_1\otimes s_2)$.\\
The minimal form factor satisfies:
$f_{ss}(\theta)=-f_{ss}(-\theta)a_{p_1}(\theta)a_{p_2}(\theta)=f_{ss}(2i\pi-\theta)$,
and reads \cite{S}:
\begin{eqnarray}
f_{ss}(\theta_{12})\equiv \frac{\cos \frac{\theta_{12}}{2i}} {\sin
\frac{\theta_{12}}{2i}}f_{p_1}(\theta_{12})f_{p_2}(\theta_{12}).
\label{ffmin}
\end{eqnarray}
It has no poles and no zeros in the physical strip
$0<\mathrm{Im}\;\theta_{12}<\pi$ and at most a simple zero at
$\theta_{12}=0$.\\
We make the following ansatz for the form factors in the $SS$
model\footnote{The numbers $s$ and $t$ of integration variables
 $u$ and $v$ are related to the topological charges $Q_1$ and $Q_2$ as well as to the number
  of particles $n$ through the formula
$Q_1=n-2s$ and $Q_2=n-2t$.}:
\begin{eqnarray}
\lefteqn{F(\theta_1,\dots,\theta_{n})=}\nonumber \\
&& \mathcal{N}_{n}\prod_{i<j}f_{ss}(\theta_{ij})\int_{C_{\theta}}du_1 \dots
\int_{C_{\theta}}du_s\; h_{p_1}(\theta,u) \Psi^{p_1}(\theta,{u})\;
\int_{C_{\theta}}dv_1 \dots \int_{C_{\theta}}dv_t\;
h_{p_2}(\theta,v)\Psi^{p_2}(\theta,{v}) \nonumber \\
&& \times \mathcal{M}_{n}(\theta,{u},{v})\; p_{n}(\theta,u,v).
 \label{ffss}
\end{eqnarray}
A few comments about our ansatz are in order:
\begin{itemize}
\item
in the equation (\ref{res}), the residue at the pole located at $\theta_1=\theta_{2n}+i\pi$ is simple.
 As explained in \cite{S}, the introduction of the cosine term in the minimal
  form factor (\ref{ffmin}) gives a zero at $\theta_1=\theta_{2n}+i\pi$, and
   consequently we should look for the residue at a second order pole of the integral
    representation proposed in equation (\ref{ff}).
\item
the properties for the $p$-function of a local operator are
similar to those of the SG model presented in the previous
section, though with a
 small modification at $\theta_1=\theta_{2n}+i\pi$:
\begin{eqnarray}
&& p_{2n}(\theta_1\dots\theta_{2n},u_1\dots u_n,v_1\dots v_n)=
-p_{2n-2}(\theta_2\dots\theta_{2n-1},u_1\dots u_{n-1},v_1\dots v_{n-1})\nonumber \\
&& +\; \tilde{p}^1(\theta_2\dots \theta_{2n-1})\quad \mathrm{at}\; u_n=v_n=\theta_{2n},\nonumber \\
&& p_{2n}(\theta_1\dots\theta_{2n},u_1\dots u_n,v_1\dots v_n)=
p_{2n-2}(\theta_2\dots\theta_{2n-1},u_1\dots u_{n-1},v_1\dots v_{n-1})\nonumber \\
&& +\; \tilde{p}^2_{\pm,\pm}(\theta_2\dots \theta_{2n-1}) \quad \mathrm{at}\;
 u_n=\theta_{2n}\pm i\pi, v_n=\theta_{2n}\pm i\pi. \nonumber
\end{eqnarray}
\item
clearly \cite{S}, the simple tensor product of the form factors of
the SG model is not a solution as it spoils the residue equation.
At $\theta_{1}=\theta_{2n}+i\pi$, each of the $2n$ integration
contours get
 pinched at $u_n=\theta_{2n},\theta_{2n}\pm i\pi$ and $v_n=\theta_{2n},\theta_{2n}\pm i\pi$
  (the choice for the integration variables $u_n$ and $v_n$ is arbitrary because of symmetry).
For this reason we introduced the function
$\mathcal{M}_{2n}(\theta,{u},{v})$ which must
   have the properties at $\theta_{1}=\theta_{2n}+i\pi$:
\begin{itemize}
\item
$
\mathcal{M}_{2n}(\theta_1\dots\theta_{2n},u_1\dots u_n,v_1\dots v_n)=\mathcal{M}_{2n-2}
(\theta_2\dots\theta_{2n-1},u_1\dots u_{n-1},v_1 \dots
v_{n-1})$
\\
when $u_n=v_n=\theta_{2n}$ or $u_n=\theta_{2n}\pm i\pi$ and
$v_n=\theta_{2n}\pm i\pi$.
\item
$ \mathcal{M}_{2n}(\theta_1 \dots \theta_{2n},u_1 \dots u_n,v_1
\dots v_n)=0 $ when $u_n=\theta_{2n}$ and $v_n=\theta_{2n}\pm
i\pi$, or when $v_n=\theta_{2n}$ and $u_n=\theta_{2n}\pm i\pi$.
\end{itemize}
\end{itemize}
We introduce the set $S=(1,\dots,2n)$ as well as $T\subset S$ and
$\bar{T}\equiv S\backslash T$. A new result of this article is
the following expression\footnote{The notation '$\#$' stands for
'number of elements'.}:
\begin{eqnarray}
\mathcal{M}_{2n}(\theta,{u},{v})=\frac{4}{\sum_{i=1}^{2n}
 e^{-\theta_i}\sum_{i=1}^{2n} e^{\theta_i}} \sum_{T \subset S, \atop \#T=n-1}
  \frac{\prod_{k,l\in \bar{T}\atop k<l}\cos^2\frac{\theta_{kl}}{2i}}
{\prod_{i\in T,\atop k\in
\bar{T}}\sin^2\frac{\theta_{ki}}{2i}} \nonumber \\
\times \; \frac{\prod_{i\in T,\atop r=1,\dots
,n}\cos\frac{\theta_i-u_r}{2i}}{\prod_{1\le r<s\le n} \cos
\frac{u_r-u_s}{2i}}\;\frac{\prod_{i\in T,\atop r=1,\dots
,n}\cos\frac{\theta_i-v_r}{2i}}{\prod_{1\le r<s\le n} \cos
\frac{v_r-v_s}{2i}}\; . \label{M}
\end{eqnarray}
For two particles, this function is equal to
$\mathcal{M}_{2}(\theta_1,\theta_2,u,v)=1$.\\
The function above is not the only one to have the required
properties, but we conjecture that this is the one we need.

\paragraph{\it Trace of the stress energy tensor.}
We propose the following $p$-function:
$$
p_{2n}^{\Theta}\left(\theta,
u,v\right)=p_{SG}^{\Theta}(\theta,u)\otimes
\left(p_{SG}^{\frac{1}{2}}(\theta,v)+p_{SG}^{-\frac{1}{2}}(\theta,v)\right)
+\left(p_{SG}^{\frac{1}{2}}(\theta,u)+ p_{SG}^{-\frac{1}{2}}(\theta,u)\right)\otimes p_{SG}^{\Theta}(\theta,v),
$$
with the recursion relation for the normalization constant:
$$
\mathcal{N}_{2n}^{\Theta}=-i\mathcal{N}_{2n-2}^{\Theta}\frac{\left(f^{min}_{p_1}(0)f^{min}_{p_2}(0)\right)^2}{8\pi
n^2}.
$$
The $p_{SG}^{\Theta}$-function is given indifferently by eq.~(\ref{trace}) or eq.~(\ref{altern}).
In the two particles case, the form factor boils down to the sum
of products
 of two form factors of the SG model, and reads explicitly:
\begin{eqnarray}
F^{\Theta}(\theta_1,\theta_2)=\frac{2i\pi
M^2}{p_1p_2}\cosh \frac{\theta_{12}}{2}\; f_{ss}(\theta_{12})\;
 \left( \frac{(s_1\otimes \bar{s}_2 + \bar{s}_1\otimes
s_2)\otimes (s_1\otimes \bar{s}_2 + \bar{s}_1\otimes
s_2)}{\sinh\frac{1}{2p_1}(i\pi-\theta_{12})\sinh\frac{1}{2p_2}(i\pi-\theta_{12})} \right).
\label{fafa}
\end{eqnarray}
It is not difficult to check that this expression satisfies
equations (\ref{rh}); this result agrees with \cite{S}. The
constant $\mathcal{N}_{2}^{\Theta}$ was chosen equal to
$\mathcal{N}_{2}^{\Theta}=
\frac{\left(\mathcal{C}_{p_1}\mathcal{C}_{p_2}\right)^4
M^2}{p_1p_2\pi }$, such that the following relation holds: $
F^{\Theta}(\theta_1+i\pi,\theta_1)= 2\pi M^2\; e_0, $ where $M$ is
the mass of the fundamental particles.\\
For $p_1=p_2=1$, $p_3=0$, we have two sine-Gordon models at the free fermion point; the expression
 of equation (\ref{fafa}) becomes as expected:
$$
F^{\Theta}(\theta_1,\theta_2)=-2\pi M^2 \sinh \frac{\theta_{12}}{2}\; e_0.
$$
A few numerical checks on the value of the central charge at the
level of two particles contributions can be found in Table 2.
Compared with the results of Table 1, one sees that the accuracy
is rather poor. We have so far no compelling explanation for this
phenomenon. It worth mentioning that in \cite{CAF}, it was already
noticed that the two-particle approximation to the correlation
function does not always give accurate results for the numerical
estimation of the central charge, in contradiction with the common
belief.
\begin{center}
\begin{table}
\begin{center}
\begin{tabular}{|c|c|c|c|}
\hline
 & $p_1$ & $c^{(2)}_{\mathrm{num}}$ & $c_{\mathrm{exact}}$\\
\hline
\hline
$p_2=1$&1&2&2\\
&2&1.9848&3\\
\hline
$p_2=2$&2&2.1586&3\\
&3&2.196&\\
&5.2&2.213&\\
&15&2.217&\\
&150&2.217&\\
\hline
$p_2$=3&3&2.255&3\\
&10&2.29&\\
\hline
$p_2=5.2$&5.3&2.3&3\\
\hline
$p_2=10$&10&2.35&3\\
\hline
$p_2=15$&15&2.36&3\\
\hline
\end{tabular}
\end{center}\caption{$SS$ model} \label{tab2}
\end{table}
\end{center}

\paragraph{\it Exponential fields.} We introduce a non locality factor
$e^{2i\pi(\pm \frac{a_1}{\alpha_1}\pm\frac{a_2}{\alpha_2})}$ in front of the product of the $S$-matrices
in the residue equation. We propose the $p$-function:
\begin{eqnarray}
\lefteqn{p_{2n}(\theta,u,v)= p_{SG}^{\frac{a_1}{\alpha_1}}(\theta,u) \otimes
p_{SG}^{\frac{a_2}{\alpha_2}+ \frac{1}{2}}(\theta,v)+
 p_{SG}^{\frac{a_1}{\alpha_1}}(\theta,u) \otimes
p_{SG}^{\frac{a_2}{\alpha_2}- \frac{1}{2}}(\theta,v)} \\
&&
+p_{SG}^{\frac{a_1}{\alpha_1}+\frac{1}{2}}(\theta,u) \otimes
 p_{SG}^{\frac{a_2}{\alpha_2}}(\theta,v)+
p_{SG}^{\frac{a_1}{\alpha_1}-\frac{1}{2}}(\theta,u) \otimes
 p_{SG}^{\frac{a_2}{\alpha_2}}(\theta,v).\nonumber
\label{durdur}
\end{eqnarray}
One notices that the following sum of two particle form factors:
$$
F^{\frac{\alpha_1}{2},\frac{\alpha_2}{2}}(\theta_{1},\theta_2)+
F^{-\frac{\alpha_1}{2},-\frac{\alpha_2}{2}}(\theta_{1},\theta_2)+
F^{-\frac{\alpha_1}{2},\frac{\alpha_2}{2}}(\theta_{1},\theta_2)+
F^{\frac{\alpha_1}{2},-\frac{\alpha_2}{2}}(\theta_{1},\theta_2)
$$
has for $p$-function\footnote{We used the fact (in particular here
for two particles) that the form factors in the SG model with the
$p$-function $p^0_{SG}$ -which corresponds to the identity
operator- are equal to 0.}
$$
(p_{SG}^1+p_{SG}^{-1})\otimes
(p_{SG}^{\frac{1}{2}}+p_{SG}^{-\frac{1}{2}})+(p_{SG}^{\frac{1}{2}}+p_{SG}^{-\frac{1}{2}})\otimes
(p_{SG}^1+p_{SG}^{-1}).
$$
Then, using the expression (\ref{altern}), we find
\begin{eqnarray}
\lefteqn{F^{\Theta}(\theta_1,\theta_2)=} \\
&& \frac{2i\pi^2 \tilde{N}_{2}^{\Theta}
}{(\mathcal{C}_{p_1}\mathcal{C}_{p_2})^4}\left(\mathrm{cotan}\frac{\pi
p_1}{2}+\mathrm{cotan}\frac{\pi p_2}{2}\right) \cosh
\frac{\theta_{12}}{2}\; f_{ss}(\theta_{12})\;
 \left( \frac{(s_1\otimes \bar{s}_2 + \bar{s}_1\otimes
s_2)\otimes (s_1\otimes \bar{s}_2 + \bar{s}_1\otimes
s_2)}{\sinh\frac{1}{2p_1}(i\pi-\theta_{12})\sinh\frac{1}{2p_2}(i\pi-\theta_{12})}
\right).\nonumber
\end{eqnarray}
This exactly reproduces the expression for the trace operator
eq.~(\ref{fafa}) above, with the new normalization constant
$$
\tilde{N}_{2}^{\Theta}=\frac{M^2(\mathcal{C}_{p_1}\mathcal{C}_{p_2})^4}{\pi\;
p_1p_2} \frac{\sin\frac{\pi p_1}{2}\sin\frac{\pi p_2}{2}}{\sin \pi
\frac{(p_1+p_2)}{2}}=
-<\Theta>\frac{(\mathcal{C}_{p_1}\mathcal{C}_{p_2})^4}{\pi^2\;
p_1p_2}.
$$
The value for the vacuum energy
\begin{eqnarray}
<\Theta>=-\pi M^2 \frac{\sin\frac{\pi p_1}{2}\sin\frac{\pi
p_2}{2}}{\sin \pi \frac{(p_1+p_2)}{2}}, \label{vev}
\end{eqnarray}
 was first
obtained by
Bethe ansatz in \cite{fateev}.\\
From the computation made above, one may conclude that the
exponential fields $\mathcal{O}_{a_1,a_2,\pm
\frac{\alpha_3}{2}}=e^{i(2a_1\varphi_1+2a_2\varphi_2\pm\alpha_3\varphi_3)}$
are associated to the $p$-function (\ref{durdur}), as the
expression of the trace operator is \cite{fateev}:
\begin{eqnarray}
\Theta \sim e^{i\alpha_3\varphi_3}\cos (\alpha_1\varphi_1+\alpha_2\varphi_2)+ e^{-i\alpha_3\varphi_3}
\cos (\alpha_1\varphi_1-\alpha_2\varphi_2).
\nonumber
\end{eqnarray}
Let us consider the case $p_1=p_2=1$ and $p_3=0$, where we have two SG models at the free fermion point. Using the results
 of \cite{L}, the expression for form factors associated to this $p$-function gives the expected result:
\begin{eqnarray}
\lefteqn{F^{a_1,a_2}_{+-,+-}(\theta_{1},\theta_2)=
\frac{\cos \frac{\theta_{12}}{2i}}
{\sin \frac{\theta_{12}}{2i}}\;
\frac{1}{\cos^2 \frac{\theta_{12}}{2i}}\times} \nonumber \\
&&
e^{\left(\sqrt{2}(a_1+a_2)+\frac{1}{2}\right)\theta_{21}}
\sin \sqrt{2}\pi (a_1+a_2)
-e^{\left(\sqrt{2}(a_1+a_2)-\frac{1}{2}\right)\theta_{21}}\sin \sqrt{2}\pi (a_1+a_2)
\nonumber\\
&& =  \frac{2i}{\cos \frac{\theta_{12}}{2i}} \sin \sqrt{2}\pi\left(a_1+a_2\right)\; e^{\sqrt{2}(a_1+a_2)\theta_{21}}.
\nonumber
\end{eqnarray}
For more than two particles, there should be some significant simplifications in our general formula.\\
Fateev and Lashkevich constructed in \cite{FL} with a very different method
the form factors for
the most general exponential fields of the type $e^{2ia_1\varphi_1+2ia_2\varphi_2+i(2a_3\pm \alpha_3)\varphi_3}$, which
 supposes that we introduce
in the $p$-function a dependence with respect to an extra
parameter $a_3$, that does not affect the non locality factor. For
the time being, I do not see how to do this for $a_3$ arbitrary.
The first proposal that one could imagine for the exponential
fields $e^{2i(a_1\varphi_1+a_2\varphi_2\pm
\frac{k}{2}\alpha_3\varphi_3)}$, where $k$ is an odd integer, is
the following $p$-function:
\begin{eqnarray}
\lefteqn{p_{2n}(\theta,u,v)= p_{SG}^{\frac{a_1}{\alpha_1}}(\theta,u) \otimes
p_{SG}^{\frac{a_2}{\alpha_2}+ \frac{k}{2}}(\theta,v)+
 p_{SG}^{\frac{a_1}{\alpha_1}}(\theta,u) \otimes
p_{SG}^{\frac{a_2}{\alpha_2}- \frac{k}{2}}(\theta,v)}\nonumber \\
&&+p_{SG}^{\frac{a_1}{\alpha_1}+\frac{k}{2}}(\theta,u) \otimes
 p_{SG}^{\frac{a_2}{\alpha_2}}(\theta,v)+
p_{SG}^{\frac{a_1}{\alpha_1}-\frac{k}{2}}(\theta,u) \otimes
 p_{SG}^{\frac{a_2}{\alpha_2}}(\theta,v)
\end{eqnarray}
However if one considers the free fermion point, one does not
recover the expected result (unless $k=1$),
so this proposal should be rejected.\\
It is unclear to me what the form factors of exponential fields
 are for $a_3\neq
\pm 2\alpha_3$. The results of \cite{FL} indicate it should be
possible to construct them, but I am not capable of making any
comparisons with the results of this paper: in particular the
authors of \cite{FL} have not succeeded in their article into
making explicit evaluations of their two-particles integrals for
the simpler cases $a_i=\frac{k_i}{2}\alpha_i$, so even their two
particle form factor for the trace operator is not explicitly
computed.\\ One may think of introducing a dependence w.r.t $a_3$
by considering linear combination of functions of the type
(\ref{M}) where we would change the number of elements of $T$.
However, such a function (\ref{M}) is very close to a similar
function obtained in closely related mathematical problems though
different physical contexts in \cite{MS,BKS}; in these articles
only the function with $\#T=n-1$ was considered, and it seems to
me that this is what should be done also here\footnote{It is not
even clear to the author whether the form factors constructed
above really correspond to the exponential fields of the UV CFT
$e^{2ia_1\varphi_1+2ia_2\varphi_2+\pm
i\alpha_3\varphi_3}$~-although we have little doubt about the
trace operator-and are not only a formal mathematical solution of
the equations. In other words, the author has no clear
understanding of the correspondence between space of states in the
CFT and space of states in the corresponding integrable perturbed
model, beside the sine-Gordon case.}.
\paragraph{\it RSOS restrictions.}
\begin{itemize}
\item
For integer values of $p_2$ and $p_2\geq 3$, the QFT admits a
quantum group restriction with respect to the symmetry group
$U_{q_2}(sl_2)$, $q_2=\exp (\frac{2i\pi}{p_2})$, giving the
parafermionic sine-Gordon models \cite{ABL} with action:
\begin{eqnarray}
\mathcal{A}_{p_1,p_2}=\mathcal{A}_{p_2}^{(0)}+\int
d^2x\left[\frac{1}{16\pi}(\partial_\mu\varphi
)^2-\kappa\left(\psi\bar{\psi}e^{i\rho\varphi}+\psi^{\dagger}\bar{\psi}^{\dagger}e^{-i\rho\varphi}\right)\right],
\label{paraf}
\end{eqnarray}
where $\mathcal{A}_{p_2}^{(0)}$ is the action of the
$\mathbb{Z}_{p_2-2}$ parafermionic CFT with central charge
$c=2-6/p_2$, and the fields $\psi,\psi^{\dagger}$,
($\bar{\psi},\bar{\psi}^{\dagger})$ are the holomorphic
(antiholomorphic) parafermionic currents with spin
$\Delta=1-\frac{1}{p_2-2}$ ($\Delta=-\bar{\Delta}$). The field
$\varphi$ is a scalar boson field and the parameter $\rho$ is
given by
$$
\rho^2=\frac{8\pi \;p_1}{(p_2-2)(p_1+p_2-2)}.
$$
We denote the QFT (\ref{paraf}) by $\mathcal{P}(p_1,p_2)$. The
perturbing operator has conformal dimension
$\Delta_{pert}=\frac{\rho^2}{8\pi} +1-\frac{1}{p_2-2}$. In
addition to the conformal symmetry, the $\mathbb{Z}_{p_2-2}$
parafermionic CFT possesses an additional symmetry generated by
the parafermionic currents $\psi$, $(\bar{\psi})$. Basic fields in
this CFT are the order parameters $\sigma_j$, $j=0,1,\dots,p_2-3$
with conformal dimension:
$$
\delta_j=\frac{j(p_2-j-2)}{2p_2(p_2-2)}.
$$
All other fields in this CFT can be obtained from the fields
$\sigma_j$ by application of the generators of the parafermionic
symmetry \cite{FZ1}. We introduce as basic operators in
$\mathcal{P}(p_1,p_2)$ the local fields:
\begin{eqnarray}
\Phi^{(j)}_{a}=\sigma_j\exp (ia \varphi). \label{field}
\end{eqnarray}
The vacuum structure, spectrum and scattering theory of the QFT
(\ref{paraf}) are described in \cite{ABL}. Its $S$-matrix is
$$
 -S_{p_1}^{SG}(\theta_{12})\otimes S_{p_2}^{RSOS}(\theta_{12}),\quad p_1\geq
 1.
$$
For $p_2=3$, $\psi\equiv 1$ and the QFT $\mathcal{P}(p_1,3)$ is
 the usual Sine-Gordon model with coupling constant $\rho^2=\beta^2= \frac{8\pi \; p_1}{1+p_1}$.\\
 For $p_2=4$, the parafermionic current $\psi(z)$ is a Majorana
 fermion and the theory $\mathcal{P}(p_1,4)$ describes the N=1
 supersymmetric sine-Gordon model with a coupling constant $\rho^2=
 \frac{8\pi \;
 p_1}{2(2+p_1)}$.\\
In order to obtain form factors of the trace of the energy
momentum tensor in this family of models
 directly from the formula for form factors of the trace in
 the $SS$-model, we use the RSOS procedure inspired from the SG
 case, and we propose as $p$-function of the operator
 $\psi\bar{\psi}e^{i\rho\varphi}+\psi^{\dagger}\bar{\psi}^{\dagger}e^{-i\rho\varphi}$:
$$
p_{2n}^{\Theta}\left(\theta, u,v\right)=
p_{SG}^{\Theta}(\theta,u)\otimes
p^{\frac{1}{2}}_{SG}(\theta,v),
$$
with the modification of the Bethe ansatz state
$\Psi^{p_2}_{\epsilon_1 \epsilon_2 \dots \epsilon_{2n}}$ like in
equation (\ref{ms}). The two-particles form factor for the trace
operator reads explicitly:
\begin{eqnarray}
\lefteqn{F^{\Theta}(\theta_1,\theta_2)=\frac{4i\pi
M^2}{p_1p_2}\cosh \frac{\theta_{12}}{2}\; f_{ss}(\theta_{12})\;
 }
\nonumber \\
 && \times \left( \frac{(s_1\otimes \bar{s}_2 + \bar{s}_1\otimes
s_2)\otimes (e^{\frac{i\pi}{2p_2}}\; s_1\otimes \bar{s}_2
+e^{-\frac{i\pi}{2p_2}}\; \bar{s}_1\otimes
s_2)}{\sinh\frac{1}{2p_1}(i\pi-\theta_{12})\sinh\frac{1}{p_2}(i\pi-\theta_{12})}\right).
\label{traceparaf}
\end{eqnarray}
It is normalized such that
\begin{eqnarray}
F^{\Theta}(\theta_1+i\pi,\theta_1)= 2\pi M^2\;(s_1\otimes
\bar{s}_2 + \bar{s}_1\otimes s_2)\otimes (e^{\frac{i\pi}{2p_2}}\;
s_1\otimes \bar{s}_2 +e^{-\frac{i\pi}{2p_2}}\; \bar{s}_1\otimes
s_2). \nonumber
\end{eqnarray}

 The expression (\ref{traceparaf}) for the two particles form factor of the
trace of the energy momentum tensor correctly reproduces in a more
or less accurate way the known value for the UV central charge
 $c=3-6/p_2$: numerical checks are presented in Table 3. One sees that increasing $p_2$ with $p_1$ fixed, the conformal
 dimension of the perturbing
 operator gets closer and closer to one, so one may expect a badly convergent integral (\ref{cth}), which would
  be responsible for the observed lack of
 precision in the numerical tests. For $p_2$ fixed and increasing $p_1$, the conformal dimension again gets close to one:
  we do not have any explanation for the fact that for $p_2\neq 3$, the accuracy becomes better for $p_1$ bigger.\\

Again, if we use the expression (\ref{altern}) for
$p_{SG}^{\Theta}$, we find the following value for the
normalization constant $\tilde{N}_2^{\Theta}$:
$$
\tilde{N}_2^{\Theta}=\frac{M^2(\mathcal{C}_{p_1}\mathcal{C}_{p_2})^4}{\pi
p_1p_2}\tan \frac{\pi p_1}{2}.
$$
It is surprising to note that had we identified the normalization
of the trace operator as we previously did:
$$
\tilde{N}_2^{\Theta}=-<\Theta>\frac{(\mathcal{C}_{p_1}\mathcal{C}_{p_2})^4}{\pi^2
p_1p_2},
$$
this would lead to the following prediction for $<\Theta>$:
$$
<\Theta>=-\pi M^2\tan \frac{\pi p_1}{2},
$$
which is certainly true for $p_2=3,5,7,\dots$, but not for
$p_2=4,6,8,\dots$, where in the latter case the v.e.v of
$<\Theta>$ should be vanishing, according to formula (\ref{vev})!
Consequently, in this case, the normalization of the two particles
form factor of the trace operator does not reproduce the correct
answer for the vacuum expectation value of the trace operator.

Although it might seem {\it a priori} most natural to propose for
$p$-function for the operators $\psi\bar{\psi}\exp ia
\varphi(x)$\footnote{We certainly expect this to be true for
$p=3$, in which case $\psi,\bar{\psi}\equiv 1$, and we should
recover the form factors for the exponential fields in the
sine-Gordon model.} (and $\psi^{\dagger}\bar{\psi}^{\dagger}\exp
ia \varphi(x)$)~:
$$
p_{2n}(\theta,u,v)= p_{SG}^{\frac{a}{\rho}}(\theta,u) \otimes
p_{SG}^{\frac{1}{2}}(\theta,v),
$$
with the modified Bethe ansatz state
$\tilde{\Psi}^{p_2}_{\epsilon_1 \epsilon_2 \dots \epsilon_{2n}}$
as defined as in equation (\ref{ms}), we are not quite certain
that this identification is correct: the problem not being a
technical one, but a problem of {\it existence} of such
operators\footnote{The author has already experienced a similar
problem within the context of the massless N=1 super sinh-Gordon
model \cite{ponsot}, where she came to the conclusion that it might be
 meaningless to consider form factors of
$e^{\alpha\varphi},\sigma e^{\alpha\varphi}$ for $\alpha$
generic.} .
\begin{center}
\begin{table}
\begin{center}
\begin{tabular}{|c|c|c|c|}
\hline
 & $p_1$ & $c^{(2)}_{\mathrm{num}}$ & $c_{\mathrm{exact}}$\\
\hline
\hline
$p_2=3$&1&1&1\\
&2&0.9924&\\
&3.6&0.9881&\\
&4&0.9878&\\
&5.2&0.9873&\\
&20&0.9869&\\
\hline
$p_2=4$&1&1.3976&1.5\\
&2&1.4404&\\
&3.6&1.4469&\\
&4&1.4473&\\
&5.2&1.4477&\\
&20&1.4480&\\
\hline
$p_2=5$&1&1.5945&1.8\\
&2&1.6819&\\
&3.6&1.7016&\\
&4&1.7029&\\
&5.2&1.7048&\\
&20&1.706&\\
\hline
$p_2=20$&1&1.9488&2.7\\
&2&2.17&\\
&3.6&2.26&\\
&4&2.27&\\
&5.2&2.29&\\
&20&2.3&\\
\hline
\end{tabular}
\end{center}\caption{Parafermionic Sine-Gordon model}\label{tab3}
\end{table}
\end{center}

\begin{center}
\begin{table}
\begin{center}
\begin{tabular}{|c|c|c|}
\hline
 $p_2$ & $c^{(2)}_{\mathrm{num}}$ & $c_{\mathrm{exact}}$\\
\hline \hline
3& 0.9924&1\\
3.5 & 1.2542&1.2857$\dots$\\
4& 1.4404&1.5\\
7.3 & 1.9422&2.1780$\dots$\\
10&2.063&2.4\\
50.3&2.211&2.8807$\dots$\\
$+\infty$&2.217 &3\\
 \hline
\end{tabular}
\end{center}\caption{N=2 supersymmetric restricted
Sine-Gordon model} \label{tab4}
\end{table}
\end{center}

Let us mention that another interesting case is obtained for
$p_1=2$ and $p_2$ arbitrary integer. The QFT $\mathcal{P}(2,p_2)$
possesses N=2 supersymmetry and is known as the N=2 supersymmetric
restricted sine-Gordon model. The UV central charge is the same as
before. The numerical checks on the central charge can be found in
Table 4. We have found interesting to present in the table above
three cases where $p_2$ is not an integer: as one may expect, the
observables depend continuously on the parameter $p_2$.

\item
For integer values of $p_1$, the previous QFT
$\mathcal{P}(p_1,p_2)$ admits an additional restriction with
respect to the second quantum group $U_{q_1}(sl_2)$,
$q_1=\exp(\frac{2i\pi}{p_1})$. The resulting QFT \cite{ABL}, which
corresponds to integrable
 perturbations with an operator $\Phi_C$ of conformal dimension $\Delta=1-\frac{2}{p_1+p_2-2}$ of
  the coset model \cite{GKO} $su(2)_{p_1-2}
\otimes su(2)_{p_2-2} /su(2)_{p_1+p_2-4}$, is denoted by
$\mathfrak{C}(p_1,p_2)$. The $S$-matrix is \cite{ABL,BR}
$$
 -S_{p_1}^{RSOS}(\theta_{12})\otimes S_{p_2}^{RSOS}(\theta_{12}),
$$
with $p_1,p_2$ integers, $p_1,p_2 \geq 3$.\\
For $p_1=3$, the QFT $\mathfrak{C}(3,p_2)$ describes minimal CFT
models perturbed by the operator $\Phi_{1,3}$ \cite{Z}.\\
For $p_1=4$, the QFT $\mathfrak{C}(4,p_2)$ describes N=1
supersymmetric minimal CFT models perturbed by the supersymmetry
preserving operator $\Phi_C$. In particular, the model
$\mathfrak{C}(4,4)$ possesses N=2 supersymmetry. The corresponding
CFT has central charge $c=1$. It describes the special point at
the critical line of the Ashkin-Teller ($\mathbb{Z}_4$) model. At
the N=2 supersymmetric point the thermal operator $\epsilon$
coincides with the operator $\Phi_C$.\\
The basic fields in the coset CFT can be represented in terms of
fields $\sigma_j$ from the $\mathbb{Z}_{p_2-2}$ CFT and a scalar
bosonic field $\varphi$ as \cite{FZ}:
\begin{eqnarray}
\Phi^{(j)}_{l,k}=\sigma_j e^{ia_{lk}\varphi}, \label{fields2}
\end{eqnarray}
where:
$$
2a_{lk}=(k-1)\rho+\frac{(1-l)}{(p_2-2)\rho}, \quad
\rho^2=\frac{8\pi \;p_1}{(p_2-2)(p_1+p_2-2)},
$$
and the integers $k<p_1$ and $l<p_1+p_2-2$ satisfy the condition
$$
|l-k-n(p_2-2)|=j;\quad n\in \mathbb{Z}_{p_2-2}.
$$
 We propose as $p$-function of the trace of the stress
energy tensor:
\begin{eqnarray}
p_{2n}^{\Theta}\left(\theta, u,v\right)=
-\left(\sum_{i=1}^{2n}e^{-\theta_i}\sum_{i=1}^{n}e^{u_i}\right)\otimes
p^{\frac{1}{2}}_{SG}(\theta,v), \label{tracersos}
\end{eqnarray}
then both Bethe ansatz states $\Psi^{p_1,p_2}$ should be modified
as in equation (\ref{ms}). Explicitly for two particles:
\begin{eqnarray}
\lefteqn{F^{\Theta}(\theta_1,\theta_2)=\frac{8i\pi
M^2}{p_1p_2}\cosh \frac{\theta_{12}}{2}\; f_{ss}(\theta_{12})\;
 }
\nonumber \\
 && \times \left( \frac{(e^{\frac{i\pi}{2p_1}}\; s_1\otimes \bar{s}_2 + e^{-\frac{i\pi}{2p_1}}\; \bar{s}_1\otimes
s_2)\otimes (e^{\frac{i\pi}{2p_2}}\; s_1\otimes \bar{s}_2
+e^{-\frac{i\pi}{2p_2}}\; \bar{s}_1\otimes
s_2)}{\sinh\frac{1}{p_1}(i\pi-\theta_{12})\sinh\frac{1}{p_2}(i\pi-\theta_{12})}\right).
\nonumber
\end{eqnarray}
It is normalized such that
$$
F^{\Theta}(\theta_1+i\pi,\theta_1)= 2\pi
M^2\;(e^{\frac{i\pi}{2p_1}}\; s_1\otimes \bar{s}_2 +
e^{-\frac{i\pi}{2p_1}}\; \bar{s}_1\otimes s_2)\otimes
(e^{\frac{i\pi}{2p_2}}\; s_1\otimes \bar{s}_2
+e^{-\frac{i\pi}{2p_2}}\; \bar{s}_1\otimes s_2).
$$
Let us note that we certainly expect for symmetry reasons the
$p$-function
\begin{eqnarray}
p_{2n}^{\Theta}\left(\theta, u,v\right)=
-p^{\frac{1}{2}}_{SG}(\theta,u)\otimes
\left(\sum_{i=1}^{2n}e^{-\theta_i}\sum_{i=1}^{n}e^{v_i}\right)
\end{eqnarray}
to lead to the same form factors as the $p$-function in eq.
(\ref{tracersos}). Also we would expect the $p$-functions
\begin{eqnarray}
p_{2n}^{\Theta}\left(\theta,
u,v\right)&=&p^{\frac{1}{2}}_{SG}(\theta,u)\otimes
p^{1}_{SG}(\theta,v),\label{tata1} \\
p_{2n}^{\Theta}\left(\theta, u,v\right)&=&
p^{1}_{SG}(\theta,u)\otimes p^{\frac{1}{2}}_{SG}(\theta,v),
\label{tata2}
\end{eqnarray}
to give the same result for the form factors.

 The numerical
checks provide results for the central charge in Table 5, to be
compared with the exact result
$c=3-6(p^{-1}_1+p^{-1}_2-(p_1+p_2-2)^{-1})$. For $p_1=3$, they are
identical to those of Table 1. Again, increasing $p_1,p_2$ make
the conformal dimension of the perturbing operator closer and
closer to one, leading to a badly converging integral (\ref{cth}),
which may be responsible for the decrease of accuracy of the
numerical results. Let us note again that the normalization
constants $\tilde{N}_2^{\Theta}$ do not reproduce the correct
v.e.v for the trace: indeed, with the $p$-function (\ref{tata1}),
we would obtain
$$
<\Theta>=-\pi M^2\tan \frac{\pi p_2}{2},
$$
and with the $p$-function (\ref{tata2})
$$
<\Theta>=-\pi M^2\tan \frac{\pi p_1}{2},
$$
whereas the correct value for $<\Theta>$ is given by (\ref{vev}).
\begin{center}
\begin{table}
\begin{center}
\begin{tabular}{|c|c|c|c|}
\hline
&$p_2$ & $c^{(2)}_{\mathrm{num}}$ & $c_{\mathrm{exact}}$\\
\hline
\hline
$p_1=3$&3&0.5&0.5\\
&4&0.6988&0.7\\
&5&0.7972&0.8\\
&20&0.9744&0.9857$\dots$\\
\hline
$p_1=4$&4&0.9924&1\\
&5&1.1429&1.1571$\dots$\\
&20&1.4268&1.4727$\dots$\\
\hline
$p_1=5$&5&1.3240&1.35\\
\hline
$p_1=20$&20&2.2&2.5578$\dots$\\
\hline
\end{tabular}
\end{center}\caption{Coset model $su(2)_{p_1-2}
\otimes su(2)_{p_2-2} /su(2)_{p_1+p_2-4}$}\label{tab5}
\end{table}
\end{center}
Certainly, if $p_2=3$ we recover the form-factors of the operators
$e^{ia_{1k}\varphi(x)}$ in the minimal model $M_{p_1}$~:
$$
p_{2n}(\theta,u,v)= p_{SG}^{\frac{a_{1k}}{\rho}}(\theta,u) \otimes
p_{SG}^{\frac{1}{2}}(\theta,v),
$$
with the modification of the two Bethe ansatz states $\Psi^{p_1,
p_2}$.
\item
A common limiting case of these two RSOS restrictions is the
Polyakov-Wiegmann model \cite{PW}, which is the same as the
perturbation of the $SU(2)$ WZNW model at level $k$ with action
$$
S=S_{\mathrm{WZNW}_k}+\lambda\int \; d^2x J^a\bar{J}^a,
$$
where $J^a$ ($\bar{J}^a$) are the holomorphic (antiholomorphic) $SU(2)$ currents. The $S$-matrix
 of the model is given by:
$$
-S^{SG}_{+\infty}(\theta_{12})\otimes S^{RSOS}_{k+2}(\theta_{12}).
$$
\begin{center}
\begin{table}
\begin{center}
\begin{tabular}{|c|c|c|}
\hline
$k+2$ & $c^{(2)}_{\mathrm{num}}$ & $c_{\mathrm{exact}}$\\
\hline
\hline
3&0.9869&1\\
4&1.4480&1.5\\
5&1.706&1.8\\
10&2.1&2.4\\
100&2.3&2.94\\
$+\infty$&2.3&3\\
\hline
\end{tabular}
\end{center}\caption{$SU(2)_k$ WZNW model}
\end{table}
\end{center}
We get the Principal Chiral Field model \cite{PW,FR} in the
extreme limit $k=+\infty$. Whether one considers
$S^{RSOS}_{+\infty}(\theta_{12})$ or
$S^{SG}_{+\infty}(\theta_{12})$ amounts to the same result.
The numerical estimations for the central charge with two particles contribution are given in Table 6.\\
The Polyakov-Wiegmann model is asymptotically free, and the trace
operator is of conformal dimension one for any $k$; from the
results above we deduce that the approximation with two particles
leads to a conformal dimension of the trace operator which is less
than one in the case of the $SU(2)$ Thirring model ($k=1$),
whereas when one increases $k$, it becomes closer and closer to
one already at the level of two particles. Numerically, the value
for the central charge is given by an integral that becomes
logarithmically divergent when the conformal dimension of the
perturbing operator is one. It is not clear why for small $k$ the
results are quite good, and poor for big $k$.
\end{itemize}

\section{Conclusion}
A few remarks should be made concerning the formula (\ref{M}),
which
 is one important result obtained in this paper.
\begin{itemize}
\item
The function $\mathcal{M}$ is universal as it does not depend on
the scattering of the theory, and appears in models where the $S$-matrix is given by a tensor
  product of different $S$-matrix. Typically, such a construction (with suitable modifications)
  is encountered in integrable massless flows \cite{MS,ponsot,GP}, which contain three different types
  of scattering, between right/right movers, left/left movers, and right/left
  movers. It has also been used in a different physical context in
  \cite{BKS}.
\item
Other mathematical solutions are obtained by changing in formula
(\ref{M})
 the number of elements of the set $T$. It is not clear to what they could correspond.
\item
 Of course it is crucial to make a comparison with the
results of \cite{FL}.
\item
The so-called one to one correspondence between operators of the
UV CFT and operators in the massive/massless theory is a issue
which deserves, to the opinion of the author, better
understanding.
\end{itemize}

\section*{Acknowledgments}
I am grateful to P.~Grinza for numerical help, V.A.~Fateev
for constructive criticism, and F.A.~Smirnov for reading a preliminary version
of this manuscript. I would also like to thank M.~Lashkevich for pointing out an omission in the
 discussion in the first electronic version and for suggesting me to consider the particular case
 of free fermion point; this really helped to improve the manuscript. Work supported by the Euclid Network
HPRN-CT-2002-00325.

\end{document}